\documentclass[aps,prl,twocolumn,showpacs,superscriptaddress]{revtex4}

\usepackage{amssymb,amsmath}
\usepackage{bm}
\usepackage{graphicx}
\usepackage[amssymb]{SIunits}

\begin{document}

\title{Low Loss Metamaterials \\
       Based on Classical Electromagnetically Induced Transparency}

\author{Ph.\@ Tassin}
\affiliation{Department of Applied Physics and Photonics, Vrije Universiteit, Brussels,
             Pleinlaan 2,B-1050 Brussels, Belgium}

\author{Lei Zhang}
\affiliation{Ames Laboratory-USDOE, and 
             Department of Physics and Astronomy, Iowa State University, Ames IA, 50011}

\author{Th.\@ Koschny}
\affiliation{Ames Laboratory-USDOE, and 
             Department of Physics and Astronomy, Iowa State University, Ames IA, 50011}
\affiliation{Institute of Electronic Structure and Lasers (IESL), FORTH, and
             Department of Material Science and Technology, University of Crete, 71110
             Heraklion, Crete, Greece.}

\author{E.N.\@ Economou}
\affiliation{Institute of Electronic Structure and Lasers (IESL), FORTH, and
             Department of Material Science and Technology, University of Crete, 71110
             Heraklion, Crete, Greece.}

\author{C.M.\@ Soukoulis}
\affiliation{Ames Laboratory-USDOE, and 
             Department of Physics and Astronomy, Iowa State University, Ames IA, 50011}
\affiliation{Institute of Electronic Structure and Lasers (IESL), FORTH, and
             Department of Material Science and Technology, University of Crete, 71110
             Heraklion, Crete, Greece.}



\date{\today}

\begin{abstract}
We demonstrate theoretically that electromagnetically induced transparency can
be achieved in metamaterials, in which electromagnetic radiation is interacting
resonantly with mesoscopic oscillators rather than with atoms. We describe
novel metamaterial designs that can support full dark resonant state upon
interaction with an electromagnetic beam and we present results of its
frequency-dependent effective permeability and permittivity.  These results,
showing a transparency window with extremely low absorption and strong
dispersion, are confirmed by accurate simulations of the electromagnetic field
propagation in the metamaterial.
\end{abstract}

\pacs{42.70.-a,42.50.Gy,41.20.Jb}

\maketitle

Electromagnetically induced transparency, often abbreviated by EIT, has been
proposed as an interface between optical quantum information used for
communication and atomic quantum states for information
storage~\cite{Harris-1997}. It is essentially a coherent process observed in
three-level atomic media, whose optical response to a laser beam is modified by
a second beam with a well-determined detuning~\cite{Harris-1990}. EIT has been
theoretically explained in terms of a dark superposition state with vanishing
excitation probability amplitude~\cite{Imamoglu-1989,Lounis-1992} or,
alternatively, by destructive quantum interference between different excitation
pathways of the excited state~\cite{Boller-1991}. Due to this process, a small
transparency window with significantly enhanced absorption length can be
observed in the frequency response of an otherwise opaque medium. At the
resonance frequency, the anomalous dispersion profile normally observed for a
two-level resonance is transformed into extremely steep normal dispersion. This
may slow down light pulses by seven orders of
magnitude~\cite{Kasapi-1995,Hau-1999}. More advanced techniques based on EIT
even allow for the storage of optical data in matter~\cite{Fleischhauer-2000}.
For a recent review, the reader can consult, e.g.,
Ref.~\cite{Fleischhauer-2005}.

EIT has been demonstrated experimentally for several media consisting of an
ensemble of non-interacting atomic systems. The first and many recent
experiments have been conducted with metal atoms in the gas
phase~\cite{Boller-1991,Liu-2001}. Later experimental work was also performed
on doped solid-state materials with optically active atomic
centers~\cite{Ham-1997,Longdell-2005} and on quantum dot based
systems~\cite{Marcinkevicius-2008}, which have longer coherence times. The
experimental handling of these setups is rather hard, since EIT is extremely
sensitive to inhomogeneous broadening, e.g., due to the Doppler effect or
quantum dot size dispersion; the setups must typically be cooled down to liquid
helium temperatures and/or magnetic fields must be
applied~\cite{Phillips-2001}.
Recently, a classical equivalent of the EIT was introduced by using coupled
optical resonators \cite{Xu-2006} and coupled planar fish-scale copper
patterns \cite{Papasimatis-2008}.

In this Letter, we demonstrate theoretically that electromagnetically induced
transparency can be achieved in metamaterials. Such
artificially fabricated materials are made up of mesoscopic structured units,
on a larger scale than the atomic, but significantly smaller than the
wavelength of the interacting radiation. This allows to characterize their
electromagnetic properties by their effective permittivity and
permeability~\cite{Smith-2006}. Metamaterials combining negative permittivity
and permeability, i.e., negative index materials (NIMs), can be obtained by
using tiny electrical circuits called split-ring resonators and continuous
wires as the metamaterial's constituent
units~\cite{Shelby-2001,Soukoulis-2006}. Metamaterials have also led to the
recently proposed optical invisibility cloaks~\cite{Pendry-2006}. The main
obstacle in realizing NIMs in the visible is dissipation in the magnetic
component due to ohmic losses. EIT in metamaterials, and its associated low
absorption, may provide a way to overcome this obstacle. In this paper, we
shall deal with the metamaterial's magnetic and electric response, which is the
real challenge of the metamaterials field.

\begin{figure}[hbt!]
\includegraphics[clip,width=7cm]{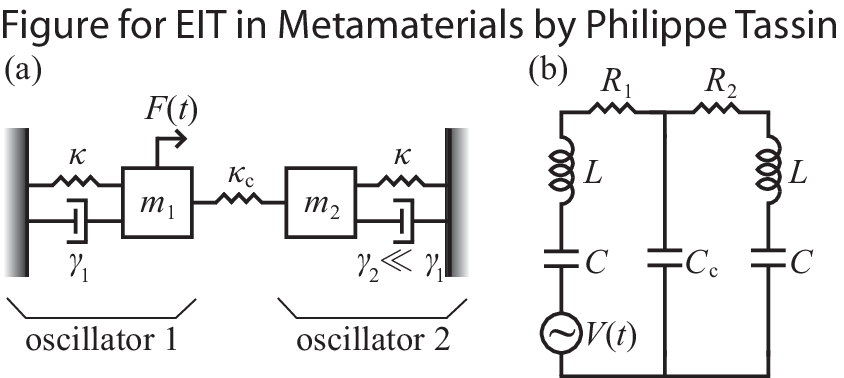}
\caption{(a) A driven mechanical system --- consisting of two damped mass-spring 
             resonators with mass $m_1 = m_2$ and spring constant $\kappa$ linearly 
             coupled by a third spring $\kappa_\mathrm{c}$ --- that exhibits 
             an EIT-like response in its power spectrum. 
         (b) Its electrical analoge.}
\label{Fig:1}
\end{figure}

It is known that EIT-like effects are in general not restricted to systems
supporting quantum mechanical states and can in principle be observed in
classical systems such as plasmas~\cite{Harris-1996} and coupled
microresonators~\cite{Smith-2004}. However, these system consist of a single
element with a pointlike input and output port, and hence cannot be considered
as a medium for propagation of electromagnetic waves. The design of the
metamaterial proposed here is based on the observation that a simple mechanical
system [see Fig.~1(a)], consisting of a driven mass-spring oscillator that is
linearly coupled to another oscillator with different dissipation factor, can
reproduce an EIT-like absorption dip in its power
spectrum~\cite{GarridoAlzar-2002}. This has been attributed to destructive
interference between the normal modes of the mechanical system at the resonance
frequency. We start by converting this mechanical model to its equivalent
electrical circuit given in Fig.~1(b): a double RLC circuit with inductance $L$
and capacitance $C$ coupled by a shared capacitor $C_c$. The two circuits have
a different resistance, respectively, $R_1$ and $R_2$. The power delivered by a
sinusoidal voltage source of magnitude $V$ to the circuit equals
\[
P = 
\mathrm{Re} 
\frac{i\omega \frac{V^2}{2}\left(L\omega^2 - i R_2\omega - \frac{1}{C_\parallel} \right)}
     {\frac{1}{C_c^2} - \left( L\omega^2 - i R_2\omega - \frac{1}{C_\parallel} \right) 
                        \left( L\omega^2 - i R_1\omega - \frac{1}{C_\parallel} \right)},
\]
with $C_\parallel = (1/C+1/C_c)^{-1}$. The above power spectrum exhibits an
EIT-like response. The dissipation minimum, which occurs approximately at the
resonance frequency, can be approximated by $P \approx (C_c^2/C_\parallel
L)(R_2 V^2/2)$ if it is assumed that the losses are small ($R_1 R_2 < L
C_\parallel/C_c^2$). It will thus suffice to make the second current loop with
low enough resistance to obtain a pronounced dissipation minimum in the power
spectrum of the circuit. We will take advantage from the fact that the second
current loop does not couple to the external field, eliminating the
contribution to $R_2$ coming from radiation losses.

\begin{figure}[hbt!]
\includegraphics[width=7cm]{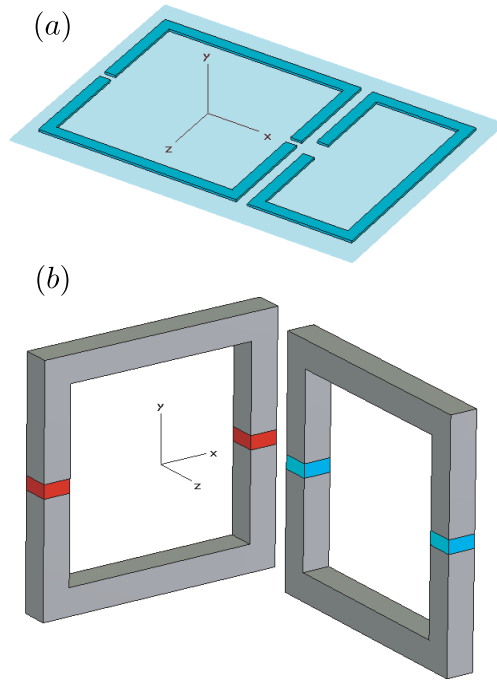}
\caption{(a) Design with electric coupling.
         The propagation direction of the incident EM wave is along the y-axis, with the electric 
         and magnetic field along the z- and x-axis, respectively.
         For this propagation direction, the resonance of the right single-gap SRR is excited electrically
         by the incident EM wave and constitutes the "radiative" or "transparent" state. 
         The dimensions are $l_1=l_2=0.9\mu m$, $w=0.1\mu m$, $g=0.0035\mu m$ for the left SRR 
         and $l_1=0.9\mu m$, $l_2=0.47\mu m$, $w=0.1\mu m$, $g=0.01\mu m$ for the right SRR.
         The distance between the two SRRs is $d=0.2\mu m$.
         (b) Design with magnetic excitation of the resonance.
         The propagation direction of the incident EM waves is along the y-axis, the electric 
         and the magnetic field is along the x- and z-axis, respectively.
         For this propagation direction, the left SRR resonance is excited by the magnetic field 
         of the incident EM radiation
         and constitutes the "radiative" or "transparent" state.
         The dimensions of the SRR are $l=0.9\mu m$, $w_1=w_2=0.1\mu m$, and $g=0.05\mu m$.
         The dielectric constants are $\varepsilon_1=10$ with loss tangent of $\tan \delta_1 = 0.01$
         and $\varepsilon_2=10$ with $\tan \delta_2 = 0.001$. 
         The distance along the z-axis between the two SRRs is $d=0.2\mu m$.
         In the simulations the SRRs are taken to be perfect metals and the losses are coming form 
         the dielectric filling in the gaps of the SRRs.
}
\label{Fig:2}
\end{figure}

The most straightforward implementation of the double RLC circuit of Fig.~1(b)
would be to etch the pattern of a ``double'' split-ring resonator as depicted
in Fig.~2(a) on printed circuit board. 
Fig.~2a describes a coupled metamaterial design that consists of a "dark" state
(left part) and a "radiative" state (right part).  For perpendicular
propagation, the EM wave couples only to the one-gap SRR on the right.  So the
"dark" state in this metamaterial is provided by the 2-gap SRR on the left.
The coupling between the "dark" and the "radiative" state is capacitive 
(and inductive, to some extent) and can be tuned by changing the distance between the two SRRs.
In contrast to atomic EIT, where the coupling between the two energy states is
realized by a pump beam, the coupling between the radiative and the dark state
in the metamaterial EIT is determined by their spatial separation.
The numerical simulations were carried out using the commercial finite
integration technique (FIT) software package CST Microwave Studio. 
For the chosen geometry of Fig.~2a, the resonance frequencyies of the isolated
SRRs are equal and around 42.9 THz.  The quality factor for the "dark" state is
equal to 120 if calculated from the frequency dependence of the absorption peak
and is equal to 800 if calculated from the frequency dependence of the
transmission dip. The corresponding values for the "radiative" SRR are 10 (from
the absorption) and 80 (from the transmission). So the quality factor of the
"dark" SRR is an order of magnitude larger than that of the "radiative" SRR.
In Fig.~3a, we present simulation results for the transmission and absorption
versus frequency.  One can clearly see a dip in the absorption, which is due to
classical EIT. 
In Fig~3b, we present the retrieval results
\cite{Smith-2006,Smith-2002,Smith-2005} for the effective dielectric
permittivity, $\varepsilon(\omega)$, which clearly shows a dip in the imaginary
part of $\varepsilon$, and one can clearly see the really strong dispersion in
the frequency dependence of $\varepsilon(\omega)$.  This strong dispersion with
simultaneously low absorption can be used to slow light for a variety of
potential applications.

\begin{figure}[hbt!]
\includegraphics[width=7cm]{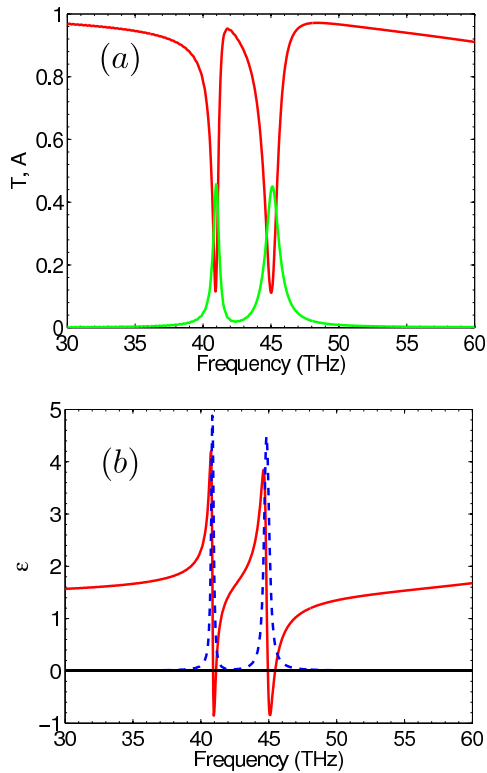}
\caption{(a) Transmission (solid line) and Absorption (dashed line) spectra for the metamaterial design in Fig.~2a.
         (b) The corresponding real (solid line) and imaginary part (dashed line) of the effective permittivity.}
\label{Fig:3}
\end{figure}

Fig.~2b describes a coupled metamaterial design that responds magnetically and
also provides a classical EIT.  This design is more difficult to fabricate
experimentally, especially at THz and optical frequencies, since it is
neccessarily three-dimensional.  The propagation direction is along the
y-direction and only one SRR can be excited magnetically, because the magnetic
field has to have a component perpendicular to the plane of the SRR in order to
couple.  So the left SRR of Fig.~2b, directly excited by the incident magnetic
field, provides the "radiative" state for the EIT.  The other SRR is the "dark"
state and can only be excited due to its  coupling to the radiative state.
For the chosen geometry of Fig.~2b, the resonance frequencies of the isolated
SRRs are equal and around 42.5THz. The quality factor for the "dark" state is
equal to 61 if calculated from the frequency dependence of the absorption peak
and is equal to 1200 if calcualted from the frequency dependence of the
transmission dip.  The corresponding values for the "radiative" SRR are 42
(from the absorption) and 160 (from the transmission).  So the quality factor
of the "dark" state is an order of magnitude larger than that of the
"radiative" SRR.
The coupling between the two SRRs is predominantly capacitive (because of the
proximity of the gaps) and can be tuned by adjusting the distance between the
two SRRs.
The coupling splits the reonance of the "radiative" SRR into two modes with 
different resonance frequencies.
At large separation between the SRRs, where the coupling is weak, the
difference between the two resonance frequencies is small, and therefore, 
the dispersion in between them is very steep. 
However, the dip in the absorption is very small. The "radiative" SRR has a 
broad absorption peak while the "dark" SRR has a narrow absorption peak. 
With increasing coupling beween the two SRRs the dip in the absorption spectrum 
widens and becomes deeper.
Simultanoeusly the dispersion becomes more flat.

\begin{figure}[hbt!]
\includegraphics[width=6cm]{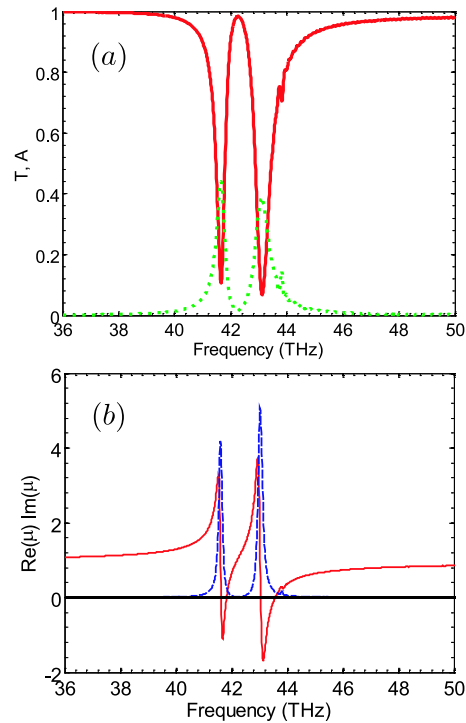}
\caption{(a) Transmission (solid line) and Absorption (dashed line) spectra for the metamaterial design in Fig.~2b.
         (b) The corresponding real (solid line) and imaginary part (dashed line) of the effective permeability.}
\label{Fig:4}
\end{figure}

In Fig.~4b, we show the retrieved real and imaginary parts of the magnetic
effective permeability for the coupled system of SRRs as shown in Fig.~2b.  One
clearly sees the classical EIT effect.  The imaginary part of $\mu$ has a very
low value at 42 THz, which signifies small loss, ocurring simultaneously with a
steep dispersion in the real part of $\mu$.  This strong dispersion gives a
very large group refractive index (order of 30), see Fig. 5, which can slow the
light for many applications.

In Fig.~5 we plot the real part of the group index, $n_g = n + \omega\,dn/d\omega$, 
versus frequency for the magnetic EIT system shown in Fig.~2b.  
Notice that $n_g$ becomes really large (of the order of 500) at the
two resonance frequencies (41.7 THz and 43.2 THz), of course the losses are
tremendous as one can see from Fig.~5.  The imaginary part of the index of
refraction is large at the two resonance frequencies, but is really small
between the frequencies 42 to 43 THz.  In this region $n_g\approx 30$ and the
imaginary part of $n$ is less than $0.01$.  So the classical EIT system, shown
in Fig.~2b, can reduce the group velocity of light by a factor of 30 with low
loss.  Similar reduction of the group velocity can be obtained for the design
shown in Fig.~2a.

\begin{figure}[hbt!]
\includegraphics[clip,width=7cm]{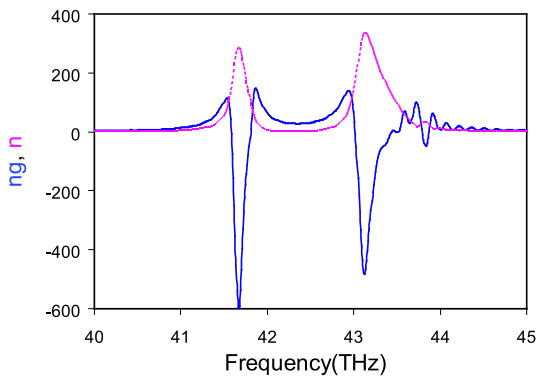}
\caption{Real part of the group index (solid line) in comparison to the imaginary part of the 
         index of refraction (dotted line) for the magnetic EIT system shown in Fig.~2b.}
\label{Fig:5}
\end{figure}

In conclusion, we have presented designs of novel metamaterials consisting of
quasistatic electric circuits with multiple independent current loops. We have
shown that these materials support a dark state leading to a phenomena similar
to electromagnetically induced transparency: the metamaterial exhibits a small
transmission window with extremely low absorption losses and steep dispersion.
No quantum mechanical atomic states are required in this metamaterial to
observe the EIT.  This could lead to ``slow light'' applications from the
microwave regime up to THz frequencies, where the structure can be most easily
fabricated. 
Furthermore, our classical EIT structures show that the effective
parameters are not always associated with high losses. therefore 
more complex designs, possibly taking advantage of knowledge from quantum
optics, could possibly provide a new way to solve the problem of losses 
in Metamaterials. Our structures already achieves a figure of merit 
$\mathrm{FOM} \approx 60$ in the transparency window for both the
electric permittivity ($\varepsilon\approx 1$) and the magnetic permeability
($\mu\approx 1$).

This work was partially supported by the FW0-Vlaanderen. P.~T.~thanks the
FWO-Vlaanderen for his ``Aspirant'' fellowship.  Work at Ames Laboratory was
supported by the Department of Energy (Basic Energy Sciences) under Contract
No.\@ DE-AC02-07CH11358.  This work was partially supported by the Office of
Naval Research (Award No.\@ N00014-07-1-D359) and European Community FET
project PHOME (Contract No.\@ 213390).

\end{document}